\documentclass[12pt,a4paper]{article}

\usepackage[dvips]{graphicx}
\usepackage{url}
\usepackage{calc}
\usepackage{times}
\usepackage{latexsym}
\usepackage{amssymb}
\usepackage{amsmath}
\usepackage{wasysym}
\usepackage[ps2pdf,hypertexnames=false,linkbordercolor={0 0 1}]{hyperref}
\usepackage{cite}
\usepackage{float}

\newcommand{\bequ}{\begin{equation}}
\newcommand{\eequ}[1]{\label{#1}\end{equation}}
\newcommand\eq[1] {(\ref{#1})}

\newcommand{\comment}[1]{}
\newcommand{\SDDT}{self-initiated DDT }

\newcommand\Le {L\!e}

\newcommand\Pra {P\!r}
\newcommand\Rey {R\!e}

\newcounter{hours}\newcounter{minutes}

\begin{document}

\DeclareGraphicsExtensions{.eps}

\title{Radiation preheating can trigger transition from 
deflagration to detonation}
\author{Vladimir Karlin\footnote{School of Forensic and 
Investigative Science, University of Central Lancashire, 
Preston PR1 2HE, UK}\ \footnote{VKarlin@uclan.ac.uk}}
\date{}
\thispagestyle{empty}
\maketitle

\begin{abstract}
In this article effect of radiation preheating of unburnt 
mixture by propagating deflagration front is studied from 
the viewpoint of its ability to form a promoting temperature 
gradient and trigger transition to detonation. First, we 
investigate the effect of radiation preheating of the 
unburnt mixtures, when they are traveling through the 
wrinkles on the flame surface, in order to estimate a 
possibility of significant temperature rise. Subsequently, 
numerical simulations of a simplified mathematical model 
are carried out. They demonstrate plausibility of the 
proposed mechanism of the deflagration to detonation 
transition.

\end{abstract}

\section{Introduction}\label{Intro}

\comment{
Technological processes in chemical industry often take
place or create explosive atmospheres of inflammable gases,
mists, vapors or dust. Not surprisingly that the principal
hazards within the chemical industry are fire and explosion.
Although systems and operating practices are designed to
prevent such catastrophes, they can occur. The gas explosion 
risk in a given situation depends on a combination of 
probability and effects of the explosion, directly related 
to explosion sensitivity and explosion severity, respectively.}

If a large amount of accidentally released gaseous fuel is 
within the limits of flammability and ignition takes place, 
a chance for transition of the deflagration front into the 
detonation wave (DDT) appears. Actualisation of the DDT 
depends on a variety of circumstances and may develop in a 
number of physical scenarios. Among all of these scenarios, 
the mechanisms not involving interference from effects, 
substances and objects external to the released fuel, i.e. 
incident shock waves, jets, particular geometry of nearby 
walls and objects, pose special interest. This 
is virtually the only possible mechanism of DDT in open 
atmosphere, and thus is especially important for safety in 
large storage and processing petrochemical plants and 
pipelines, see e.g. \cite{Buncefield06}. 

Research in explosion safety usually assumes that principles 
of safety by design rule out powerful sources of ignition in 
explosion prone industrial environment leaving chances to 
more delicate mechanisms of the DDT. For example obstacles 
placed in the way of the deflagration front or incident shock 
waves, see e.g. \cite{Oran-Gamezo07,Ciccarelli-Dorofeev08}. 
Alternatively, they can be self-initiated. 

Probably the first successful attempt to explain possible 
mechanism of \SDDT was the idea of the promoting 
temperature gradient. According to 
\cite{Zeldovich-Librovich-Makhviladze-Sivashinsky70}, 
temperature gradient in the reacting media results in a 
gradient of induction times of chemical reactions. If the 
gradient of the latter matches the speed of propagation of 
the reaction front, then the synergy of the two processes 
may transform the flame into a detonation wave. This 
scenario was carefully investigated in a number of later 
studies, see e.g. \cite{Kapila-Schwendeman-Quirk-Hawa02}, 
where an idealized medium with simple, rate-sensitive 
kinetics and an initially prescribed linear gradient of 
temperature was studied exhaustively. 

\comment{
Succession of physical phenomena leading to detonation formation 
depends on the steepness of the temperature gradient. Shallow 
gradients lead to a decelerating supersonic reaction wave with 
only weak intervention from hydrodynamics. Given enough time, 
this weak detonation wave slows down to the Chapman-Jouguet 
speed and undergoes a swift transition to the ZND structure in 
accord with the scenario suggested in 
\cite{Zeldovich-Librovich-Makhviladze-Sivashinsky70}. With 
steepening of the initial temperature gradient, gas dynamic 
nonlinearity plays a much stronger role and eventually the 
transition scenario completely changes to the one advocated in 
\cite{Lee-Moen80}. An accelerating compression pulse runs now 
ahead of the reaction wave and radically rearranges the 
distribution of induction times there. As the pulse advances, 
interaction between its leading shock, an induction zone, and 
a following fast deflagration front develop and emerge as a ZND 
detonation. Further steepening of the initial temperature 
gradient results in compression pulses which are unable to 
rearrange the initial distribution of induction times properly, 
the deflagration front lags behind, and DDT no longer takes 
place.} 

Once it was understood that temperature nonuniformities may 
induce transition to detonation, an investigation of possible 
physical mechanisms leading to such uniformities was 
undertaken. Amongst theoretical achievements in the area we 
would mention the concept of the equivalent drag force, 
which works to create temperature nonuniformities ahead of 
the flame front \cite{Sivashinsky02}. Numerical simulations 
demonstrating importanve of the effect of viscous friction in 
the boundary layer \cite{Kagan-Sivashinsky03} were carried 
out in relatively narrow channels. Although, further work 
\cite{Kagan-Sivashinsky08} indicates that the effect remains 
relevant in quite realistic cases too. 

Another approach is based on the collective thermal effect of 
shock waves generated in front of a very fast flame by the 
flame itself \cite{Liberman-Kuznetsov-Ivanov-Matsukov09}. The 
theory looks feasible for flames, which are naturally very 
fast and, in addition, experience immense exponential 
acceleration. Both computational and experimental examples 
given in \cite{Liberman-Kuznetsov-Ivanov-Matsukov09} 
demonstrate possibility of such exponential acceleration of 
flames in tubes up to supersonic propagation speeds with Mach 
numbers well in excess of one. At such speeds supersonic 
deflagration fronts are able to generate downstream shocks 
strong enough to change thermal and reactive properties of 
the fuel cardinally. It is not clear how to extend these 
ideas to unconfined outward propagating flames.

\subsection{Acceleration of expanding flames}\label{IntroAcceleration}

It was discovered in \cite{Gostintsev-Istratov-Shulenin88}, 
that expanding flames do accelerate. This acceleration was 
linked to appearance of wrinkles, as wrinkles increase 
flame surface area allowing to burn larger quantities of 
fuel and thus expand faster. Theoretical and numerical 
analysis, see e.g. 
\cite{Karlin02a,Karlin-Sivashinsky05b,Karlin-Sivashinsky05a}, 
revealed physical mechanism and basic characteristics of the 
acceleration. It was established that flame fronts become 
extremely sensitive to the upstream velocity perturbations 
as their size grows. From the physical point of view, the 
sensitivity of cellular flames to perturbations of the 
upstream flow velocity is explained by the possibility of 
a pseudoresonant interaction between them \cite{Karlin05a}. 
The pseudoresonant interaction amplifies certain upstream 
velocity perturbations and transforms them into pairs of 
coupled vortexes of size proportional to the the thermal 
flame thickness, which are located just behind the flame 
front. These pairs of vortexes suck the unburnt gas into 
the areas occupied by the products of combustion and form 
wrinkles which significantly increase the bulk burning 
rate and hence the averaged flame propagation speed.

The power law of growth of the averaged flame radius 
$\bar{r}(t)\propto(t-t_{*})^{\beta}$ for large enough times 
$t>t_{*}$, in accordance with the fractal analysis  
\cite{Liberman-Ivanov-Peil-Valiev-Eriksson04}, was confirmed. 
Values of the power $\beta$ depend on the dimension of the 
problem and equal $3/2$ for three dimensional flames and $5/4$ 
for two-dimensional ones. In expanding flames, the 
pseudoresonant interaction between the flame and the upstream 
flow velocity perturbations is weakened by stretch. As a 
result set up of the acceleration at $t_{*}$ can be delayed 
significantly. On the other hand, permanent growth of the size 
of the flame results in its permanent acceleration unlike 
the non-stretched planar flames, which have a finite limit 
on their propagation speed \cite{Karlin-Sivashinsky05a}.
Using a realistic set of dimensional parameters, e.g. planar
flame speed relative to the burnt gases $u_{b}=0.5\,$m/s
and thermal diffusivity $D_{th}=2.5\times 10^{-5}\,$m$^{2}$/s,
our findings for $t_{*}$ in dimensional terms can be put as 
$\bar{r}(t_{*})\approx 20\,$m.

Theoretically, continuous acceleration of the flame should 
lead to supersonic speeds and further to DDT. However, the 
rate of expansion of spherical flames 
$\bar{r}(t)\propto(t-t_{*})^{3/2}$, is so slow that even 
an optimistic estimation, e.g. assuming initial flame expansion 
rate of about $v_{0}\approx 66\,$m/s for flame of averaged 
radius $r_{0}\approx 20\,$m, gives flame radius $\bar{r}$ 
of order of $r_{0}(v/v_{0})^{3}\approx 2.5\,$km for which 
supersonic speeds $v\approx 330\,$m/s can be reached. This 
means that if self-initiated DDT of expanding flames takes 
place at all, it should be assisted, and even more likely 
dominated, by an effect other than flame acceleration.

\subsection{Radiation preheating}\label{IntroPreheating}

Upon realizing that flame acceleration due to flame 
cellularization alone cannot lead to self-initiated DDT, 
appearance of a promoting temperature gradient in front of 
expanding flames looks the only factor which can. The 
question is how such temperature gradient can be generated 
in front of the expanding flame? Consideration of time 
scales specific to heat transfer via conduction and 
radiation, shows that the most likely heat transfer mode 
which is able to modify the temperature gradient is 
radiation. Normally, gaseous fuels do not absorb radiative 
heat significantly, however even minute quantities of water 
vapour can change this behaviour dramatically. Moreover, 
fine aerosol and dust fuels often are good radiation heat 
absorbents. Initiation of detonation by ultraviolet 
radiation was demonstrated in 
\cite{Lee-Knystautas-Yoshikawa78}, where the radiation 
was an external factor and its direct action was 
photochemical rather than the thermal one. 

\begin{figure}[ht]
\begin{center}
\begin{minipage}{45mm}
\centerline{\includegraphics[height=40mm,width=40mm]{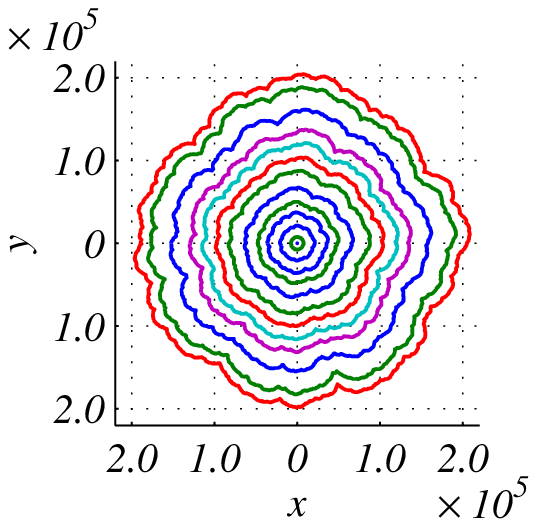}}
\end{minipage}\begin{minipage}{65mm}
\centerline{\includegraphics[width=60mm,height=33mm]{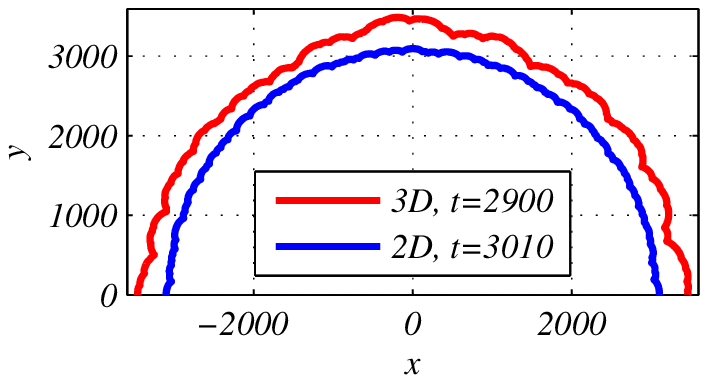}}
\end{minipage}
\caption{Development of a fractal structure on surface of circular 
flame (left). Comparison of the circular flame and the equatorial 
cross-section of the three-dimensional spherical flame at 
nearly the same time instances (right).}\label{TwoThreeComp3}
\end{center}
\end{figure}

Calculations show that the depth of the wrinkles on 
two-dimensional circular flame surface reaches up to 1/10-th 
of the averaged flame radius. Comparison with the 
three-dimensional spherical flames shows that the 
latter expand and grow cellular structures even faster, see 
Fig. \ref{TwoThreeComp3} from \cite{Karlin-Sivashinsky05b}. 
Their cellular structures are better 
developed and wrinkles are even deeper. This suggests that 
some unburnt fluid particles may spend significant time 
inside the wrinkle before reaching the flame front itself. 
Thus, preheating of the unburnt mixture, traveling inside 
the wrinkles, with the heat radiation from surrounding 
flame surfaces might be important.

In this work we carry out tentative estimations of a possibility 
of the effect of radiation preheating to trigger the DDT. In 
Section \ref{Estimations} we consider a simple analytical model of the 
phenomena confirming a possibility of significant growth of 
temperature due to radiation in deep enough wrinkles. Further, in 
Section \ref{Numerics}, a Navier-Stokes based numerical model is 
presented. Results of a set of computer simulations of the 
phenomena in question are discussed in Section \ref{Results}, 
indicating that the DDT in combustible mixtures with essential 
heat emissivity and absorption is possible. Physical mechanism 
of such heat radiation triggered \SDDT\!'s is explained too.

\section{Preliminary estimations}\label{Estimations}

\subsection{Heat balance equation}\label{EstimationsEquation}
\begin{figure}[ht]
\centerline{\includegraphics[width=75mm,height=86mm]{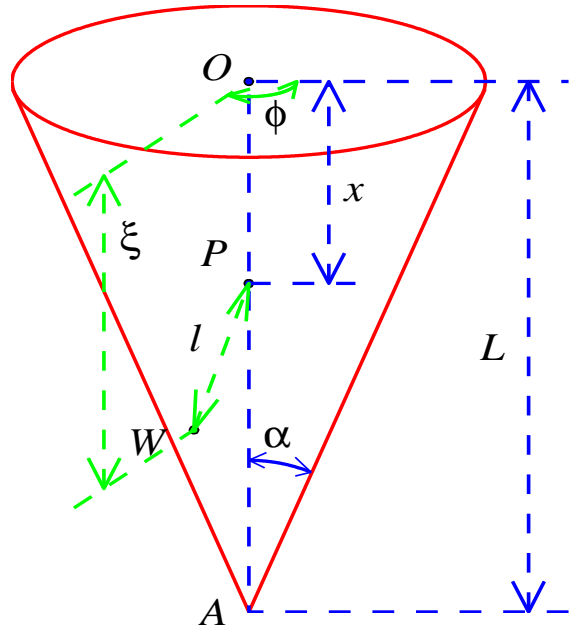}}
\caption{Model of a wrinkle on the surface of an expanding 
flame.}\label{WrinkleModel}
\end{figure}

Let us assume that a combustible fluid particle $P$ is moving 
inside a flame wrinkle and estimate its temperature growth 
because of heat radiated from hot flame. The estimation will 
be by order of amplitude, similar to that one outlined in 
\cite{Karlin07b}. For the sake of simplicity the products of 
combustion are assumed at the same uniform temperature 
$T_{W}$ and particle $P$ is not very close to the flame yet. 
Under this conditions, the only heat which matters is the heat 
from the flame absorbed by the particle. All other elements of 
the radiating system are in nearly thermodynamical equilibrium 
and radiation heat exchange between 
them can be neglected. Furthermore, the wrinkle will be modelled 
as a conical cavern of depth $L$ and of the apex semi-angle $\alpha$, 
see Fig. \ref{WrinkleModel}. Under assumptions in question, 
variations of the geometry of the wrinkle result just in a 
coefficient of order one and are not essential. 

Every element $W$ of area $dA_{W}$ of the cone surface radiates 
heat according to the Stefan-Boltzmann law with the coefficient 
of emissivity $\varepsilon_{W}$. So that heat radiated per unit time is
$\left.\dfrac{dE}{dt}\right|_{W}
=\varepsilon_{W}\sigma T_{W}^{4}dA_{W}$.
The heat is transferred from $W$ to $P$, which is at a distance $l$, 
dissipating according to the exponential law with the dissipation 
factor $\chi$. It means that heat radiated from point $W$ per 
unit time will be weakened by $e^{-\rho_{a}\chi l}$ times upon travelling 
through distance $l$. Here $\rho_{a}$ is the density of the 
absorbing species, which might be quite different from the 
averaged density $\rho$ of the combustible mixture. 

The heat flux incident on the fluid particle $P$ of cross-section 
area $dA_{P}$ is the areal fraction $dA_{P}/(2\pi l^{2})$ of 
the radiation energy which reaches $P$ after being radiated from 
$W$ is $\left.\dfrac{dE}{dt}\right|_{W\rightarrow P}
=\dfrac{dA_{P}}{2\pi l^{2}}e^{-\rho_{a}\chi l}
\varepsilon_{W}\sigma T_{W}^{4}dA_{W}$.
Hence, the fraction of this radiation heat which will dissipate 
within the small fluid particle $P$ of thickness $dl$ is
$\left.\dfrac{dE}{dt}\right|_{P\leftarrow W}=\dfrac{dA_{P}}{2\pi l^{2}}
\left(\rho_{a}\chi dl\right)e^{-\rho_{a}\chi l}
\left(\varepsilon_{W}\sigma T_{W}^{4}\right)dA_{W}$.
Denoting volume of the fluid particle $P$ as $dV=dA_{P}dl$, 
the total rate of heat gain of the fluid particle $P$ is
\[
\frac{dE_{P}}{dVdt}
=\frac{\varepsilon_{W}\sigma\rho_{a}\chi T_{W}^{4}}{2\pi}
\int\limits_{A_{W}}l^{-2}e^{-\rho_{a}\chi l}dA_{W}.
\]

Noticing that the $x,\xi$-averaged 
$l^{2}=(\xi-x)^{2}+(\xi-L)^{2}(\tan\alpha)^{2}$ is proportional 
to $L^{2}$, $dA_{W}=\dfrac{L-\xi}{\cos\alpha}d\xi d\phi$, 
and that the integrand of the above integral is positive, 
the integral itself can be approximated as follows
\bequ
\frac{dE_{P}}{dVdt}
\approx \frac{\varepsilon_{W}\sigma\rho_{a}\chi T_{W}^{4}
e^{-\rho_{a}\chi L\sqrt{c_{g}}}}{2c_{g}\cos\alpha}.
\eequ{RadGain}
where $c_{g}$ is a ``geometry correction coefficient'' of 
order $O(1)$.

As temperature of the fluid particle grows due to radiation 
heating, intensity of chemical reaction starts to grow too.
We estimate the rate of this heat release assuming an overall 
exothermic uni-molecular irreversible chemical reaction 
$\mathcal{A}\rightarrow\mathcal{B}$. For example, $\mathcal{A}$ 
is a mixture of $CH_{4}$ and air and $\mathcal{B}$ is a 
mixture of air, $CO_{2}$, and gaseous $H_{2}O$. If the 
reaction produces energy $Q$ per mole of reacted species 
$\mathcal{A}$, then the heat release because of chemical 
reactions in the fluid particle per unit volume per 
unit time is 
$\rho\dfrac{dc_{P}T}{dt}=Q\rho^{\nu}k_{a}e^{-\frac{E_{a}}{RT}}$,
where $c_{P}$ is the specific heat capacity at constant 
pressure, $\nu$ is the order of the reaction, $k_{a}$ is the 
pre-exponential factor, and $E_{a}$ is the activation energy.
Assuming that the fluid particle is moving toward the flame 
front with velocity $u_{n}$, we can replace time $t$ with 
$x=u_{n}t$:
\bequ
c_{P}\rho u_{n}\frac{dT}{dx}=Q\rho^{\nu}k_{a}e^{-\frac{E_{a}}{RT}},
\eequ{ChemGain}

Combining heats produced by radiative heat flux from the 
flame surface \eq{RadGain} and chemical reactions \eq{ChemGain}: 
\bequ
\frac{d\Theta}{d\zeta}
\approx q_{c}e^{\frac{T_{a}\Theta}{1+\Theta}}+q_{r}, 
\eequ{TotalGain}
where 
$q_{c}=\dfrac{Q\rho^{\nu-1}k_{a}Le^{-T_{a}}}{c_{P}u_{n}T_{0}}$,
$q_{r}=\dfrac{\varepsilon_{W}\sigma T_{W}^{4}\rho_{a}\chi L
e^{-\rho_{a}\chi L\sqrt{c_{g}}}}{c_{g}\rho c_{P}u_{n}T_{0}\cos\alpha}$,
$T_{a}=\dfrac{E_{a}}{RT_{0}}$, $\Theta=\dfrac{T-T_{0}}{T_{0}}$, 
and $\zeta=x/L$.

\subsection{Approximate solution}\label{EstimationsSolution}

In the early stages of heating, fluid particle temperature 
does not differ from $T_{0}$ significantly and the Arrhenius 
exponent can be approximated as
$\exp\left(\dfrac{T_{a}\Theta}{1+\Theta}\right)\approx e^{T_{a}\Theta}$.
Then, the correspondingly approximated heat balance equation 
\eq{TotalGain} is integrated exactly
$\Theta(\zeta)\approx-\dfrac{1}{T_{a}}\ln
\left[\left(1+\dfrac{q_{c}}{q_{r}}\right)e^{-q_{r}T_{a}\zeta}
-\dfrac{q_{c}}{q_{r}}\right]$, yielding the solution blowup 
distance for $\zeta=x/L$ equal
\bequ
\zeta^{*}\approx
\frac{1}{q_{r}T_{a}}\ln\left(1+\frac{q_{r}}{q_{c}}\right).
\eequ{TotalGainSol}
It is interesting to compare the latter to the blowup distance 
$\zeta_{0}^{*}\approx\dfrac{1}{q_{c}T_{a}}$ for $q_{r}\equiv 0$.

Considering parameters typical to combustion of hydrocarbons 
in air $T_{a}=80$, $Q/(c_{P}T_{0})=8$, $u_{n}=0.5\,$m/s, 
$k_{a}=5\times 10^{12}\,$s$^{-1}$, $\nu=1$, one obtains
$q_{c}\approx 2\times 10^{-22}L$. Taking into account that 
function $f(x)=xe^{-x}$ reaches its maximum equal $e^{-1}$ 
for $x=1$, we conclude that $q_{r}$ is maximal for 
$\rho_{a}\chi L=c_{g}^{-1/2}$ and this maximum is
\bequ
q_{r,\max}\approx\dfrac{\varepsilon_{W}}{ec_{g}^{3/2}\cos\alpha}
\times\dfrac{\sigma T_{W}^{4}}{\rho c_{P}u_{n}T_{0}}.
\eequ{qrmax}
Assuming that the first factor in \eq{qrmax} is about $1/2$, 
$\rho=1\,$kg/m$^{3}$ and $T_{W}=1500\,$K, one obtains 
from \eq{TotalGainSol} that $\zeta^{*}\approx 0.9$ for 
$L\gtrsim 1\,$m. Numerical integration of \eq{TotalGain} shows 
that significant growth of temperatures can be reached for 
even smaller flames. According to our earlier analysis, 
wrinkles of depth $1\,$m can appear on flames of radius 
$r_{c}\gtrapprox 10\,$m.

Our choice of $\rho_{a}\chi L=c_{g}^{-1/2}$ may look artificial. 
However, it is not, because the value of the product is 
effectively controlled by the density $\rho_{a}$ of the heat 
absorbing admixture and by the wrinkle depth $L$. The former, is at 
the disposal of the experimentalist. For some particular species 
values of $\rho_{a}$ suitable for efficient combustion might 
be limited and the limit may lie below the 
requirements of the efficient radiation heat absorption for 
given $\chi$ and $L$. Even so, required values of $\rho_{a}$ 
can be always lowered below this limit for deeper wrinkles, 
i.e. for larger $L$. 

These simplified estimations were carried out for some 
``typical'' values of combustion and heat transfer parameters 
and for an optimal radiation heat absorption. Thus, they 
just show that radiation preheating of fuel in front of the 
deflagration flame might be significant in principle. Such 
temperature rise alone not necessarily leads to the DDT yet 
and more detailed numerical simulations are still required in 
order to validate the hypothesis. Furthermore, mixtures with 
such a combination of the parameters in question might be 
difficult to find in practice even if theoretical numerical 
simulations confirm that radiation preheating can cause the 
DDT. 

As we have mentioned in Section \ref{IntroPreheating}, two 
heat transfer mechanisms potentially able to produce 
promoting temperature gradient are heat conduction
and radiation. According to \eq{qrmax}, the time scale of 
the radiation preheating over distances of order 
$l_{a}=(\rho_{a}\chi)^{-1}$ is of order $\tau_{r}
=\dfrac{\rho c_{P}l_{a}}{\varepsilon_{W}\sigma T_{W}^{3}}$. 
In contrast, the time scale of heat transfer by conduction 
over the same distance is $\tau_{c}=l_{a}^{2}/\kappa$, where 
$\kappa$ is the thermal diffusivity. By representing 
$\kappa\propto\delta_{th}u_{n}$ in terms of the thermal 
flame thickness $\delta_{th}$ and laminar burning velocity 
$u_{n}$, the ratio of the time scales takes the form
\[
\frac{\tau_{r}}{\tau_{c}}
=\frac{\rho c_{P}u_{n}}{\sigma T_{W}^{3}}
\times\frac{\delta_{th}}{l_{a}}\varepsilon_{W}^{-1}.
\]

For gases value of $\rho c_{P}$ is typically about 
$10^{3}\,$J/(m$^{3}\times$K) and $u_{n}$ is of order 
$1\,$m/s. Then, value of the Stefan-Boltzmann constant 
$\sigma\approx 5.67\times 10^{-8}\,$kg/(s$^{3}\times$K$^{4}$) 
and reasonable flame temperatures of at least 
$T_{W}=1800\,$K ensure that the first factor in 
\eq{TimeScales} does not exceed 3. Hence, for typical 
gaseous fuels
\bequ
\frac{\tau_{r}}{\tau_{c}}
=\frac{3}{\varepsilon_{W}}
\times\frac{\delta_{th}}{l_{a}}.
\eequ{TimeScales}

Thickness of the preheated layer $l_{T}$, which renders 
optimal temperature gradient, exceeds $\delta_{th}$ by a 
few orders of magnitude \cite{Kapila-Schwendeman-Quirk-Hawa02}. 
Thus, for reasonably radiating flames, i.e. 
$\varepsilon_{W}>0.03$, and if we manage to adjust $\rho_{a}$ 
to make the dissipation length $l_{a}$ match $l_{T}$, 
radiation heat transfer will dominate conductivity. 

Relationship between the leading time scales is getting 
more complicated for multi-component and especially for 
multiphase fuels. For example, if radiation absorbing and 
reacting species are different, then the speed of heat 
transfer between them should be taken into account too. 
Further complications are linked to the time scales 
of evaporation and pyrolysis processes, which are relevant 
to burning of mist and dust mixtures.

\section{Mathematical model and its numerical implementation}\label{Numerics}

Numerical simulations of dynamics of deflagration fronts 
in presence of intensive radiation preheating was studied 
in the case of planar flames propagating in tubes open from 
one end rather than for expanding spherical flames in the 
open. This geometrical simplification made it possible to 
carry out investigations with relatively limited resources. 

The approach used Navier-Stokes description of compressible 
fluid and employed drastically simplified chemistry and 
radiation models. The governing equations can be written 
as follows
\[
\frac{\partial V}{\partial t}
+\frac{\partial\mathcal{F}(V)}{\partial x_{1}}
+\frac{\partial\mathcal{G}(V)}{\partial x_{2}}
=\frac{\partial\mathcal{F}^{(d)}(V)}{\partial x_{1}}
+\frac{\partial\mathcal{G}^{(d)}(V)}{\partial x_{2}}
+\mathcal{Q}(V),
\]
where $V=\left[\rho,\rho u_{1},\rho u_{2},\frac{\rho T}{\gamma-1}
+\frac{\gamma M^{2}}{2}\rho\left(u_{1}^{2}+u_{2}^{2}\right),\rho Y\right]$.
Formulas for convective $\mathcal{F}(V)$, $\mathcal{G}(V)$, and 
dissipative $\mathcal{F}^{(d)}(V)$, $\mathcal{G}^{(d)}(V)$ fluxes 
can be found elsewhere. The source term $\mathcal{Q}(V)$ is combined 
of contributions from chemistry and radiation. Tube diameter $d$ 
and laminar burning speed relevant to the burnt gases $u_{b}$ 
were used as scales, so that the Mach number $M=\dfrac{u_{b}}{a_{0}}$, 
where $a_{0}$ is the speed of sound in the fuel at initial 
temperature $T_{0}$. Other nondimensional governing parameters 
used were the Reynolds number $\Rey=\dfrac{\rho_{0}u_{b}d}{\mu}$, 
$T_{a}=\dfrac{E_{a}}{RT_{0}}$, and $Q=\dfrac{\bar{Q}}{c_{v}T_{0}}$.
Prandtl $\Pra=\dfrac{c_{p}\mu}{\kappa}$ and Lewis 
$\Le=\dfrac{\kappa}{c_{p}D\rho_{0}}$ numbers were set to one.

Chemistry was modelled with a single irreversible 
$\mathcal{A}\rightarrow\mathcal{B}$ reaction of Arrhenius type 
similar to the assumptions in Section \ref{EstimationsEquation}. 
The radiation model was reduced to the heat flux driven by 
temperature according to the Stefan-Boltzmann law. 
In addition, emission by cold gas and gas too far away 
from the flame as well as absorption by hot gas and gas too 
far away from the flame were not taken into account, because 
gases in those areas are in nearly thermodynamical equilibrium 
from the view point of radiation. In spite of its roughness, 
the model governs interaction of all relevant energy fluxes 
reasonably well.

The governing equations of the model were solved with a high 
order upwind shock capturing finite difference scheme without 
any fractional steps. Numerical approximation is implicit in 
the stream-wise coordinate, and is explicit in the other one, 
similar to \cite{Lyubimov-Rusanov70}, providing satisfactory 
compromise between limitations of numerical stability and 
efficient parallelization. Computational domain was moving 
together with the flame front and absorbing boundary 
conditions were implemented on its up- and downstream 
boundaries. The resulting code is reasonably efficient for 
Mach numbers greater than 0.01, Reynolds numbers less than 
1000, and for the nondimensional activation energy $T_{a}$ 
less than 50.

\section{Computer simulations and discussion}\label{Results}

\subsection{Numerical experiments}\label{ResultsExperiments}
\begin{figure}[ht]
\centerline{\includegraphics[width=30mm,height=30mm]{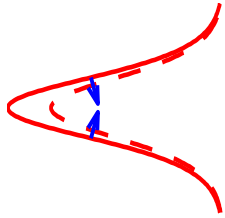}}
\caption{Formation of an optimal temperature gradient by 
radiation preheating.}\label{Tgrad}
\end{figure}

\begin{figure}[ht]
\centerline{\includegraphics[width=72mm,height=90mm]{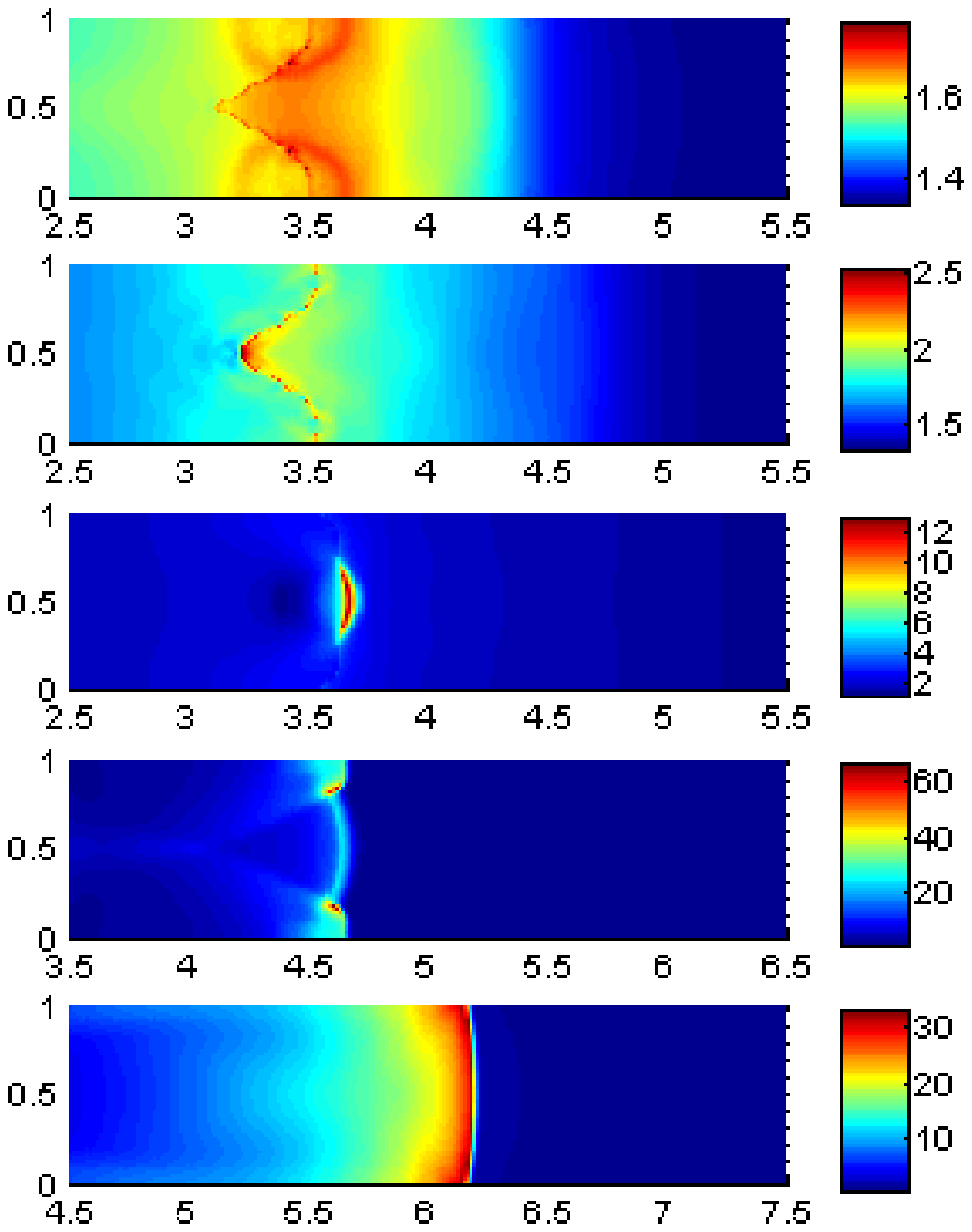}
\includegraphics[width=72mm,height=90mm]{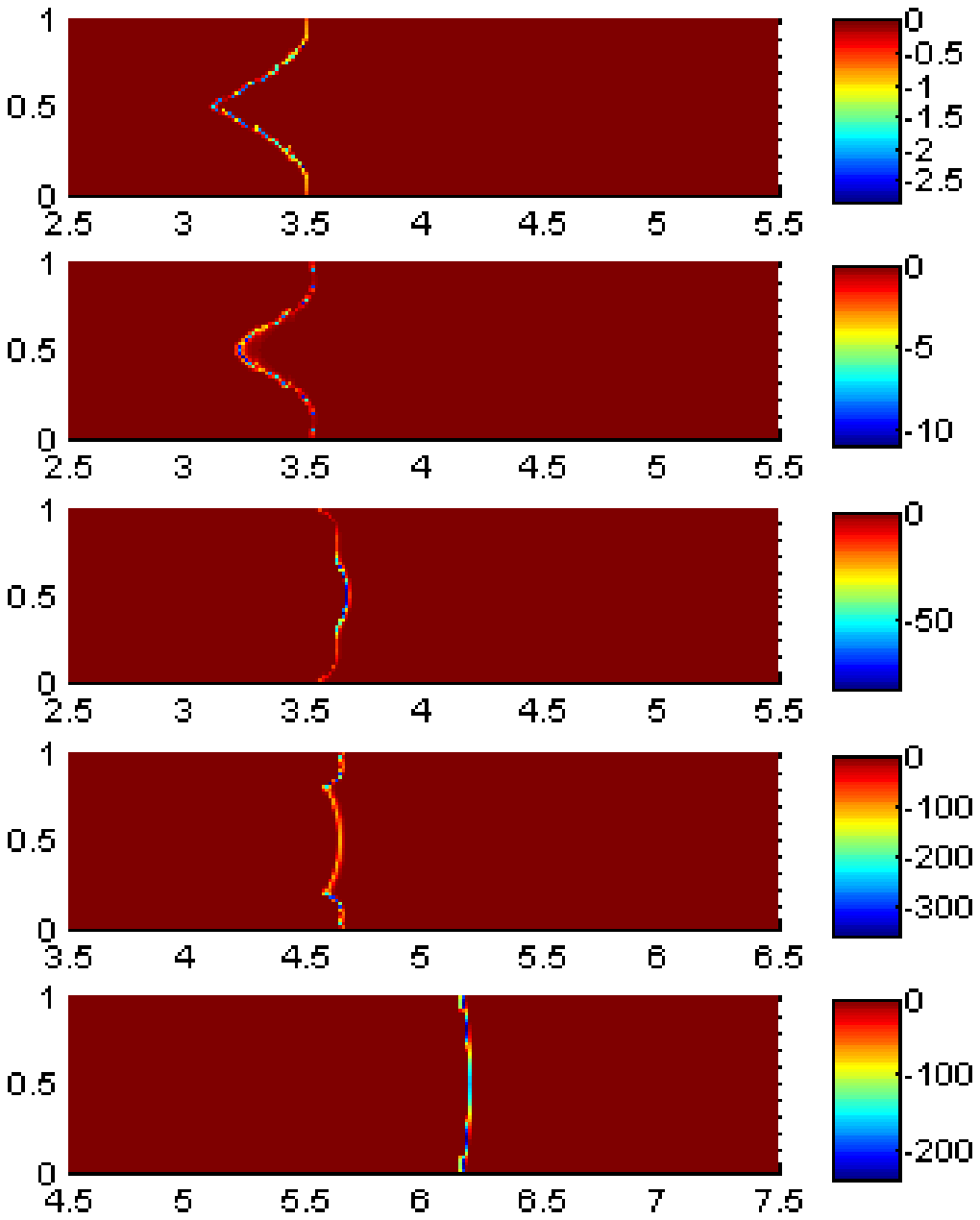}}
\caption{Example of evolution of pressure (left) and 
reaction rate (right) fields in the self-initiation 
of DDT.}\label{Simulation2}
\end{figure}

In our numerical simulations self-initiation of DDT never 
took place in smooth open pipes without obstacles and 
absorption of radiation heat. On the other hand, significant 
absorption of radiation heat was able to alter behavior of 
the flame front dramatically. It was found that 
self-initiation of DDT in radiation heat absorbing media 
progresses via formation of an optimal temperature gradient 
on the sides of wrinkles on the flame surface. Successful 
DDT was resulting from interaction (``collision'') of two 
flame segments accelerating through such promoting 
temperature gradients as illustrated in Fig \ref{Tgrad}. 
Zone of intensive chemical reaction is indicated with a 
solid line in there, while the boundary of the layer of 
unburnt gas at essentially elevated temperature is shown 
with a dashed line. Normals to the flame surface indicate 
presence of parts of the preheated layer in which 
structure of the temperature gradient is the most 
advantageous for the deflagration front to transform into 
a detonation wave. Appearance of such a thick preheated 
layer of the unburnt gas can be seen in the second row 
in Fig. \ref{Simulation2} as well. 

Sufficient amount of fuel available between two opposing 
flame segments of the wrinkle is critical for success of 
the DDT. In addition, the effect is sensitive to 
the shape of the flame front, which in turn depends on a 
set of governing parameters of the problem. For example, 
chemical reactions of orders higher than one widen range 
of suitable flames able to experience the DDT significantly, 
confirming earlier observations of other researchers, see 
e.g. \cite{Kagan-Liberman-Sivashinsky07}. Note, heat 
radiation from the flame not only heats up the unburnt fuel 
in front, but cools down the flame itself too, which levels 
out the gradient even faster.

In contrast to expectations based on common sense, success 
of the \SDDT depends on the depth of the flame wrinkle in a 
lesser degree, which makes our estimation of critical flame 
radius of order $r_{c}\approx 10\,$m in Section 
\ref{EstimationsSolution} less relevant. The \SDDT based on 
the mechanism just described might work for much smaller 
flames. An asymptotic estimation of the critical radius of 
outward propagating unconfined flames for this new \SDDT 
mechanism is possible, but would deserve a separate 
investigation. 

\begin{figure}[ht]
\centerline{\includegraphics[width=72mm,height=72mm]{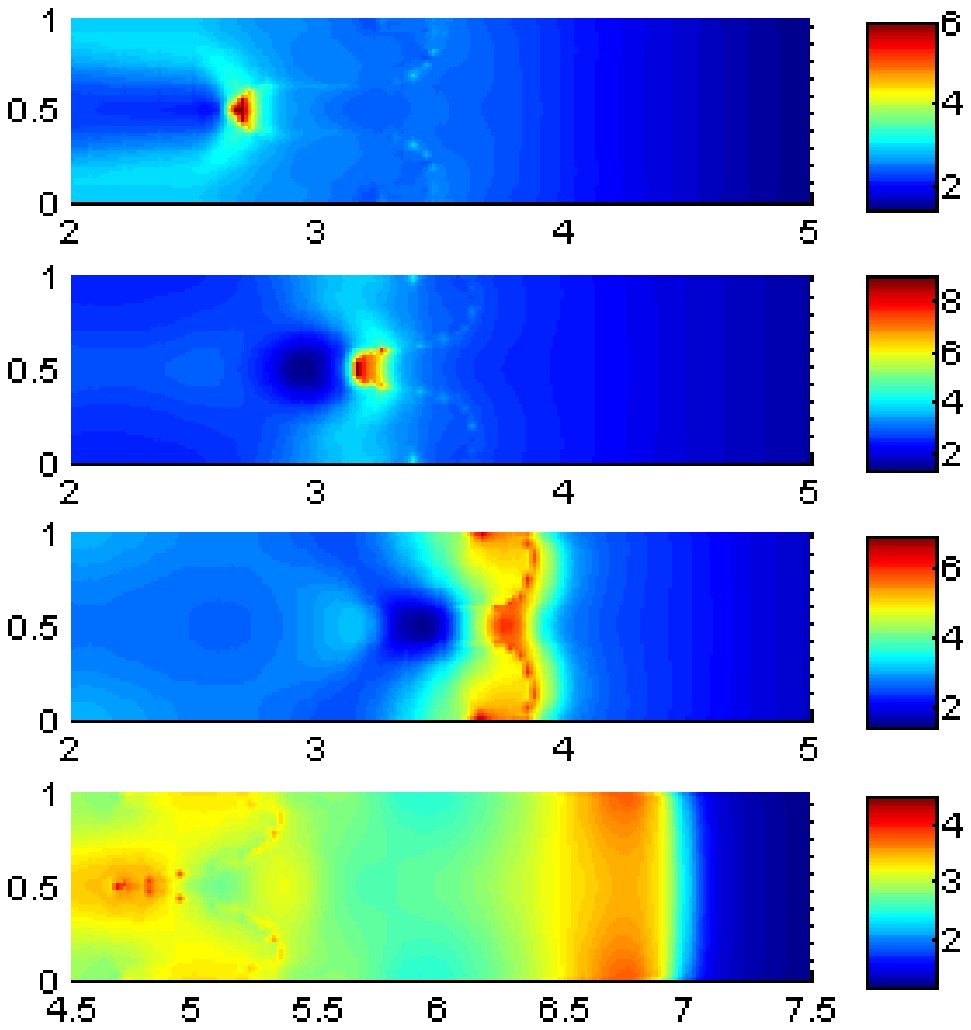}
\includegraphics[width=72mm,height=72mm]{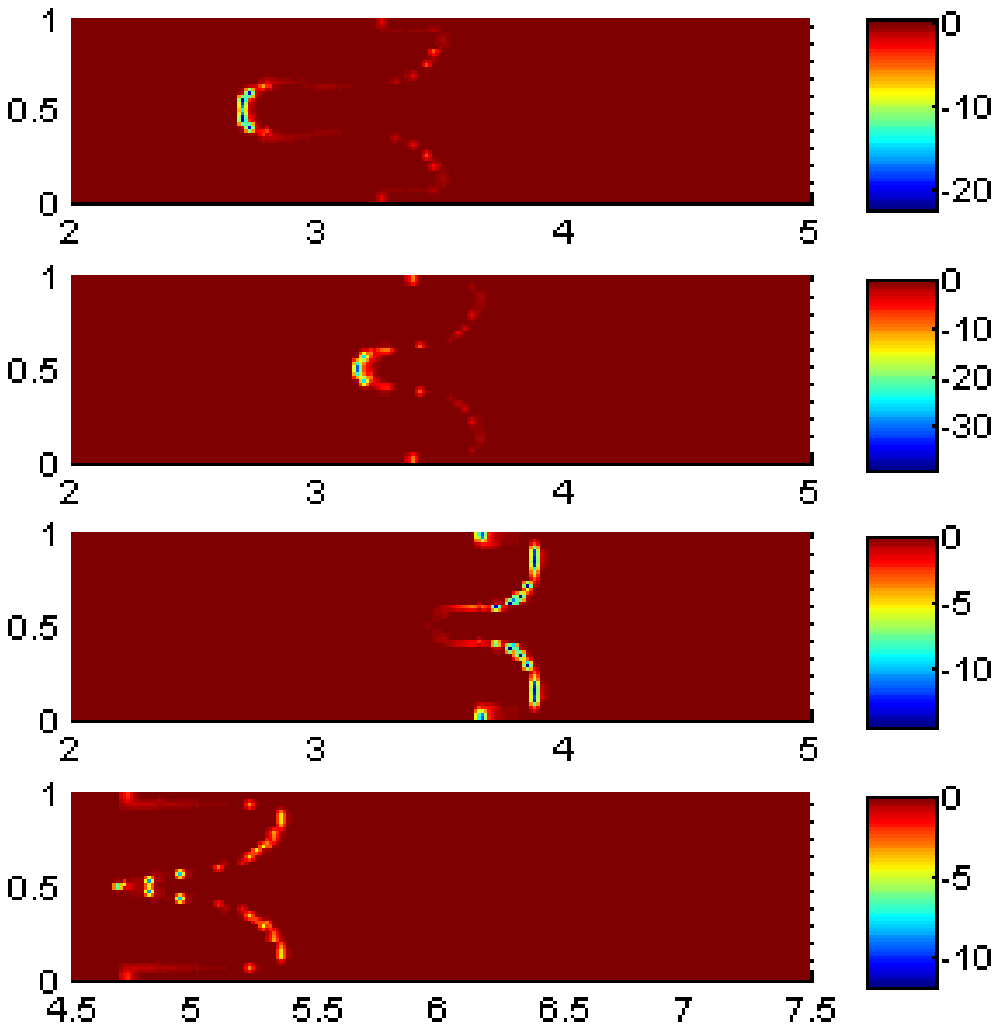}}
\caption{Example of evolution of pressure (left) and reaction 
rate (right) fields in the failed self-initiation of DDT.}\label{Simulation1}
\end{figure}

As absorption rate grows, flame dynamics evolves through a 
sequence of distinctive regimes. First, the flame is not 
affected by the radiation at all. Rise of the absorption 
rate leads to formation of promoting temperature gradients 
able to generate powerful compression waves. However, 
these temperature gradients are insufficient to allow for 
proper acceleration of deflagration fronts yet and 
generated compression waves are not strong enough to 
maintain supersonic combustion behind them. As a result, 
the steady flame propagation is altered by a sequence of 
failed detonations as illustrated in Fig. \ref{Simulation1}.

Eventually, when absorption rate is high enough, stable 
\SDDT takes place \cite{Karlin09} as this is illustrated 
in Fig \ref{Simulation2}. The first row depicts the flame 
just before the DDT. The next row illustrates formation 
of the preheated layer of the unburnt gas in front of the 
flame. Soon after events begin to unfold really fast. The 
third row shows the moment when a ``micro-explosion'' inside 
the wrinkle just took place. Generated powerful compression 
wave kick starts the detonation front shown in the following 
rows. Reminiscences of the shock waves brought forth by the 
opposing flame segments continue to interfere with the 
downstream propagating detonation front for some time as 
can be seen in the forth row of Fig. \ref{Simulation2}. 
However, they decay eventually and the detonation wave 
settles gradually to a perfect ZND detonation structure 
exemplified in the fifth row.

Effect of radiation heat transfer at very high radiation 
absorption rate was not studied yet, but is believed to 
be similar to the case of high thermal conductivity. 
Hence, dynamics of the \SDDT at very high radiation 
absorption rate is expected to be similar to the case 
when optimal temperature gradient is formed by means of 
heat conductivity rather than by the radiation heat 
transfer \eq{TimeScales}. It is interesting to note that 
the detonation initiations presented in 
\cite{Kagan-Liberman-Sivashinsky07} were obtained for 
rather high thermal conductivity and that the range of 
flame wrinkles, able to generate a detonation, widens as 
heat conductivity grows. All this can be explained in 
terms of formation of an optimal temperature gradient by 
means of conduction. It is also known that at high 
thermal conductivity flames lose ability to form 
properly shaped wrinkles and an artificial corrugated 
wall was needed in \cite{Kagan-Liberman-Sivashinsky07} 
to create them. Introduction of strong fuel diffusion, 
i.e. consideration of small Lewis numbers, see e.g. 
\cite{Karlin-etal00}, might fix the problem too.

\subsection{Critical analysis}\label{ResultsAnalysis}

Practical realization of the \SDDT mechanism in question is 
expected for mixtures containing water vapour, fine liquid 
sprays and dusts because variation of concentration of the 
admixtures affects values of the dissipation factor $\rho_{a}\chi$ in 
a very wide range. Many experimentalists report that moist 
atmospheres increase chances of DDT significantly, see e.g. 
\cite{Ennis08}. However, in practice, choice of mixtures 
with required properties might be limited. Moreover, mixtures 
with specific characteristics, which can be easily considered 
theoretically, may simply not exist in real at all. Hence, 
validation of practical usefulness of the proposed mechanism 
of DDT requires development of realistic chemical and heat 
radiation/absorption models. These models should be coupled, 
i.e. changes in the concentration of water vapours 
should affect not only rates of heat radiation and absorption 
in the mixture, but rates of chemical reactions as well. High 
enough concentration of water vapours will result in almost as 
high heat radiation/absorption rate as required, but at certain 
stage their excess will begin to slow down or even completely 
extinguish the flame. 

On the other hand, once quite detailed chemistry is going to 
be considered, the heat radiation/absorption model has to be 
refined too. As concentration of water vapour 
changes, new intermediate species with special radiation 
properties may appear. Therefore, the heat radiation/absorption 
model should take into account the spectral structure of 
the radiation fluxes rather than their total energy alone. 
Furthermore, if heat absorbing and reacting species are not 
the same, then heat exchange between them should be taken into 
account too.

Relatively low Reynolds numbers and activation energies were 
used in our numerical simulations. Character of changes as 
the Reynolds number grows does not suggest any problems up 
to transition to turbulence. The latter looks just a technical 
challenge rather than the undoer of the proposed DDT scenario. 
Changes in the character of flame dynamics when activation 
energy grows suggest just further shrinking of the range of 
absorption rate for which detonation repeatedly fails. The 
tendency explains lack of experimental observations of the 
regime of successively failing detonations. 

Eventually, this investigation is based on a simplified 
geometrical model of the phenomena. The original problem of
\SDDT is a three-dimensional one and is in the open atmosphere, 
whereas only a planar two-dimensional combustion in a pipe was 
considered in this paper.

\section{Conclusions}
Possible temperature rise of gas fuel in front of propagating
flame as a result of radiation heat transfer was estimated. 
The estimation demonstrated that temperature rise of fuel
approaching flame inside a wrinkle can match the adiabatic 
burning temperature for relatively small flames of radii just 
$10\,$m as soon as the gas mixtures involved are able to emit 
and absorb radiation heat at high enough rate. 

A variety of numerical experiments on flame dynamics in pipes 
open from one end was carried out as well. As absorption rate 
grows, flame dynamics evolves through a sequence of regimes. 
Initially flame is not affected by the radiation. At certain 
rate of radiation heat absorption flame propagation is altered 
by a sequence of failed detonations. Eventually, stable 
\SDDT takes place. 

Undertaken simulations demonstrate plausibility of the idea of 
the radiation preheating as the principal effect in the \SDDT. 
However, the \SDDT is a very delicate phenomena. More accurate 
and detailed numerical simulations are needed in order to 
clarify issues highlighted in this work.

\section*{Acknowledgements}

The author appreciates a possibility to use the National HPC 
HECToR service under the auspices of the UK Consortium on 
Computational Combustion for Engineering Applications (COCCFEA), 
EPSRC grant EP/D080223.

\bibliography{arXiv_DDT_VK}

\end{document}